\documentclass[sigconf]{acmart}

\copyrightyear{2026}
\acmYear{2026}
\setcopyright{cc}
\setcctype{by-nc-nd}
\acmConference[MM '26]{Proceedings of the 34th ACM International Conference on Multimedia}{November 10--14, 2026}{Rio de Janeiro, Brazil}
\acmBooktitle{Proceedings of the 34th ACM International Conference on Multimedia (MM '26), November 10--14, 2026, Rio de Janeiro, Brazil}
\acmDOI{10.1145/3767308.3836630}
\acmISBN{979-8-4007-2213-4/2026/11}

\usepackage{booktabs}
\usepackage{graphicx}
\usepackage{balance}

\graphicspath{{figures/}}

\AtBeginDocument{%
  }

\begin{document}

\title{Hidden-Domain Routing for All-Type Audio Deepfake Detection}

\author{Yifan Gao}
\email{gaoyifan@oppo.com}
\affiliation{%
  \institution{OPPO}
  \department{AI Center}
  \city{Beijing}
  \country{China}
}

\author{Yao Tian}
\email{aaron1@oppo.com}
\affiliation{%
  \institution{OPPO}
  \department{AI Center}
  \city{Beijing}
  \country{China}
}

\author{Hongbin Suo}
\email{suohongbin@oppo.com}
\affiliation{%
  \institution{OPPO}
  \department{AI Center}
  \city{Beijing}
  \country{China}
}

\author{Haonan Lu}
\correspondingauthor
\email{luhaonan@oppo.com}
\affiliation{%
  \institution{OPPO}
  \department{AI Center}
  \city{Shenzhen}
  \country{China}
}

\renewcommand{\shortauthors}{Yifan Gao, Yao Tian, Hongbin Suo, and Haonan Lu}

\begin{abstract}
All-type audio deepfake detection requires authenticity decisions across
speech, environmental sound, singing voice, and music, while the audio
type is unavailable at inference time. In AT-ADD Track2, this setting
creates a hidden audio-domain condition: the binary real/fake label is
shared across domains, but representation structure and detector-score
behavior vary with audio type. We present a closed-condition routed
system that first recovers the hidden audio domain and then interprets
detector scores within the selected branch. The AudioType-BEATs-6s
Router estimates audio type from a 6-second window; speech inputs are
handled by the Speech-XLSR Expert, while sound, singing, and music rely
on EAT-based general-audio experts with branch-local score
interpretation. Development-set representation analysis, router-family
comparisons, and component results show audio-domain separation and
complementary detector strengths across audio types. On the official
AT-ADD Track2 final evaluation, the system achieves 96.10\% Track2
Macro-F1 and ranks first on the final leaderboard, with type-wise
Macro-F1 scores of 88.07\%, 98.18\%, 99.07\%, and 99.08\% for speech,
sound, singing, and music, respectively. These results support
recovering the hidden audio domain before interpreting detector scores
in all-type audio deepfake detection.
\end{abstract}




\maketitle

\section{Introduction}

Recent neural audio generation models can produce increasingly realistic
speech, environmental sound, singing voice, and music, extending audio
deepfake detection from speech spoofing to a broader all-type setting
\cite{whatsreal2024,svdd2024,esdd2026,fakemusiccaps2024}.
Text-to-speech and voice conversion have long motivated speech spoofing
countermeasures \cite{asvspoof2021,add2022}. More recent
AI-synthesized voice and audio language model studies further show that
fake-audio detection must cover heterogeneous acoustic content rather
than only spoken utterances \cite{whatsreal2024,allm4add2025}. This
shift raises a practical robustness question: whether a detector can
maintain reliable authenticity decisions as the audio content type
changes.

Shared evaluation campaigns have played a central role in defining audio
deepfake detection tasks. The ASVspoof series established standardized
speech spoofing benchmarks for synthesized, converted, replayed, and
deepfake speech \cite{asvspoof2019,asvspoof2021}. The ADD challenges
further broadened speech-oriented audio deepfake detection to include
audio deep synthesis, low-quality media, and in-the-wild conditions
\cite{add2022,add2023}. Recent task-specific benchmarks and datasets
have also studied singing voice deepfake detection
\cite{svdd2024,singfake2024}, environmental sound deepfake detection
\cite{esdd2026}, and synthetic music detection
\cite{fakemusiccaps2024}. These efforts show that fake-audio detection
now spans multiple content domains, although many protocols still
evaluate one domain at a time.

The All-Type Audio Deepfake Detection (AT-ADD) challenge extends these
efforts to a unified heterogeneous-audio benchmark \cite{atadd2026}. In
Track2, each system is required to submit a binary real/fake prediction
for an input clip that may contain speech, environmental sound, singing
voice, or music. The audio type is not provided at inference time. For
each audio type, the official evaluation computes the Macro-F1 over the
real and fake classes, and the final Track2 score is the average of the
four type-wise Macro-F1 values. The task therefore combines a shared
binary authenticity label with a hidden audio-domain condition.

This protocol induces a domain-conditioned detection problem: the audio
type is hidden at inference time, but it affects the acoustic structure
of the input. Speech, environmental sound, singing voice, and music
differ in source structure and temporal organization, and generated
samples in these domains can lead to different detector-relevant cues.
These domains also favor different representation priors:
speech-oriented self-supervised models such as wav2vec 2.0
\cite{wav2vec2} and XLS-R \cite{xlsr2022} are optimized for spoken
utterances, whereas general-audio representations such as BEATs
\cite{beats2023} and EAT \cite{eat2024} target broader acoustic events.
As a result, a detector score can have different decision semantics
across domains even when the final label space is binary. We therefore
treat AT-ADD Track2 as binary detection under a hidden audio-domain
condition.

Based on this formulation, we build a routed system under the
closed-track setting. The AudioType-BEATs-6s Router first predicts the
hidden audio type from a 6-second window. The predicted type selects one
detector branch, so speech inputs are evaluated by a speech detector
combining XLS-R and AASIST \cite{xlsr2022,aasist2022}, while non-speech
inputs are evaluated by EAT-based general-audio detectors
\cite{eat2024}. Each branch then applies a branch-local decision rule
for the recovered domain. This design makes audio type an explicit
condition for score interpretation.

The submitted routed system achieves 96.10\% Track2 Macro-F1 on the
official AT-ADD Track2 final evaluation and ranks first on the
leaderboard. The type-wise Macro-F1 scores are 88.07\%, 98.18\%,
99.07\%, and 99.08\% for speech, sound, singing, and music,
respectively.

The main contributions are summarized as follows:
\begin{itemize}
  \item We formulate all-type audio deepfake detection as binary
  detection under a hidden audio-domain condition, where audio type
  affects representation geometry and detector-score behavior.
  \item We develop a closed-track routed system that first recovers the
  hidden audio type and then applies branch-specific detector scores
  with branch-local decision rules.
  \item We provide empirical evidence from development-set
  representation analysis, router-family comparison, component results,
  and official final aggregate scores, achieving Rank 1 on AT-ADD
  Track2 with 96.10\% Track2 Macro-F1.
\end{itemize}

\begin{figure*}[t!]
  \centering
  \includegraphics[width=\textwidth]{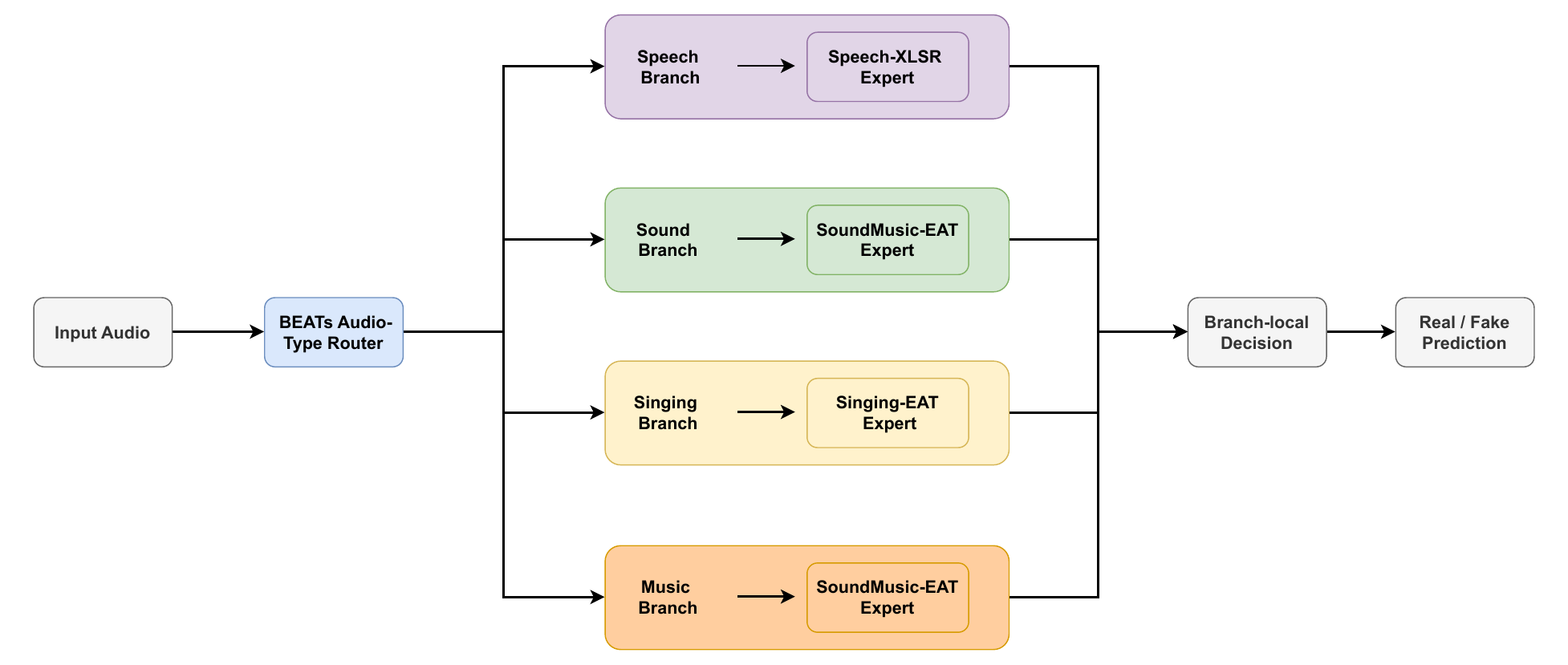}
  \caption{Routed all-type audio deepfake detection pipeline. The
  AudioType-BEATs-6s Router predicts the hidden audio type, and the
  selected branch applies its detector-score set and branch-local
  decision rule.}
  \Description{A system diagram. Input audio is passed to the
  AudioType-BEATs-6s Router. The router activates exactly one of four
  branches: speech, sound, singing, or music. The speech branch uses the
  Speech-XLSR Expert, the sound and music branches use the
  SoundMusic-EAT Expert, and the singing branch uses the Singing-EAT
  Expert. The selected branch is passed to a branch-local decision
  module, which outputs the final real or fake prediction.}
  \label{fig:system_overview}
\end{figure*}

\section{Related Work}

\subsection{Speech-Centric Audio Deepfake Detection}

Audio deepfake detection has been studied primarily in speech-centric
settings. ASVspoof \cite{asvspoof2019,asvspoof2021} and ADD
\cite{add2022,add2023} established influential benchmarks covering
synthesized, converted, replayed, low-quality, partially fake, and
in-the-wild speech. Earlier countermeasures often relied on handcrafted
spectral features and lightweight classifiers. Raw-waveform frontends
and augmentation methods, including RawNet-style models
\cite{rawnet2019} and RawBoost \cite{rawboost2022}, brought
waveform-level modeling into anti-spoofing pipelines. Neural spoofing
backends such as AASIST \cite{aasist2022} and SSL frontends for spoofing
detection \cite{wav2vec2spoof2022} then became common high-performing
components in speech spoofing countermeasures. Recent studies on XLS-R
layer selection \cite{xlsrsls2024}, SSL feature gating
\cite{sslgating2025}, and audio LLM-based detectors
\cite{allm4add2025} have also reported strong performance in
speech-oriented ADD.

This line of work has largely concentrated on speech-domain spoofing and
deepfake detection, with representations and backends designed around
spoken utterances. All-type evaluation extends this scope by evaluating
speech together with environmental sound, singing voice, and music while
the type label is unavailable at inference time.

\begin{table*}[t!]
  \centering
  \caption{Official AT-ADD Track2 training and development split
  statistics. For each audio type, Clips is the sample count and Mean is
  the mean duration in seconds.}
  \label{tab:track2_distribution}
  \begin{tabular}{@{}lrrrrrrrrrrr@{}}
    \toprule
    Split & \multicolumn{1}{c}{Total} & \multicolumn{1}{c}{Real} &
    \multicolumn{1}{c}{Fake} & \multicolumn{2}{c}{Speech} &
    \multicolumn{2}{c}{Sound} & \multicolumn{2}{c}{Singing} &
    \multicolumn{2}{c}{Music} \\
    \cmidrule(lr){5-6} \cmidrule(lr){7-8}
    \cmidrule(lr){9-10} \cmidrule(lr){11-12}
    & & & & \multicolumn{1}{c}{Clips} & \multicolumn{1}{c}{Mean} &
    \multicolumn{1}{c}{Clips} & \multicolumn{1}{c}{Mean} &
    \multicolumn{1}{c}{Clips} & \multicolumn{1}{c}{Mean} &
    \multicolumn{1}{c}{Clips} & \multicolumn{1}{c}{Mean} \\
    \midrule
    Train & 146,781 & 33,150 & 113,631 & 49,575 & 5.23 &
    39,840 & 8.72 & 36,000 & 5.28 & 21,366 & 10.04 \\
    Dev & 91,069 & 19,476 & 71,593 & 49,734 & 5.34 &
    19,929 & 8.72 & 16,000 & 4.94 & 5,406 & 10.04 \\
    \bottomrule
  \end{tabular}
\end{table*}

\subsection{All-Type and Non-Speech Audio Deepfake Detection}

Recent tasks and benchmarks have expanded audio deepfake detection
beyond spoken utterances. Work on singing voice deepfake detection
examines singing-specific generation artifacts and cross-singer or
cross-domain generalization \cite{svdd2024,singfake2024}.
Environmental sound deepfake detection focuses on generated or
manipulated acoustic events with structures different from speech
\cite{esdd2026,envsdd2025}. Synthetic music detection and attribution
require detectors to handle long-range musical structure, timbre, and
text-to-music generation settings \cite{fakemusiccaps2024}. These
tasks indicate that detection cues vary across audio domains.

AT-ADD \cite{atadd2026} combines speech, environmental sound, singing
voice, and music under one binary prediction protocol, with the audio
type hidden at inference time. The official splits also exhibit
substantial differences in sample counts and mean durations across
types (Table~\ref{tab:track2_distribution}). WPT-SSL
\cite{wptssl2026} further studies detection in this multi-domain
setting.

\subsection{Self-Supervised Audio Representations and Countermeasures}

Self-supervised learning has become central to audio deepfake
countermeasures. Speech SSL models such as wav2vec 2.0
\cite{wav2vec2}, XLS-R \cite{xlsr2022}, HuBERT \cite{hubert2021},
WavLM \cite{wavlm2022}, and W2V-BERT \cite{w2vbert2021} learn from
large speech corpora and are widely used with spoofing backends such as
AASIST \cite{aasist2022}.

General-audio representations instead learn from broader acoustic
content. AST \cite{ast2021}, AudioMAE \cite{audiomae2022}, BEATs
\cite{beats2023}, EAT \cite{eat2024}, and CLAP \cite{clap2023}
provide transformer-based, masked-prediction, acoustic-tokenization, or
audio-language pretraining families. This distinction motivates using
different representation priors for audio-type recovery and
domain-conditioned authenticity detection.

\subsection{Domain Conditioning, Routing, and Score Interpretation}

All-type audio deepfake detection is related to dataset shift
\cite{datasetshift2009} and domain generalization. Methods such as DANN
\cite{dann2016}, IRM \cite{irm2019}, and DomainBed
\cite{domainbed2021} seek predictors that remain reliable under domain
changes. Audio deepfake studies have likewise examined cross-domain
generalization and whether failures arise from genuinely different
distributions rather than uniformly harder samples
\cite{asdg2024,harderdifferent2024,crossdomainadd2024}.

Routing offers an explicit way to condition decisions on a latent or
observed domain. Mixture-of-experts models use input-dependent selection
to allocate samples to specialized components
\cite{localexperts1991,hierarchicalmoe1994,sparsemoe2017}. Classifier
calibration work further shows that scores can become miscalibrated
across distributions \cite{calibration2017,labelshift2018}. We
therefore treat the recovered audio domain as a condition for both
expert selection and score interpretation.

\section{Method}

\subsection{System Overview}

In AT-ADD Track2, systems predict whether an input clip is real or fake
across four acoustically different domains: speech, environmental sound,
singing voice, and music. Because the audio type is unavailable at
inference time, we formulate the task as binary detection under a hidden
audio-domain condition rather than assuming one shared score space for
all inputs.

Given an input clip \(x\), the audio-type router estimates the hidden
domain
\[
\hat{z}=\arg\max_{z \in \mathcal{Z}} R_z(x), \quad
\mathcal{Z}=\{\mathrm{speech}, \mathrm{sound}, \mathrm{singing},
\mathrm{music}\},
\]
where \(R_z(x)\) is the router probability for type \(z\). The top-1
type selects exactly one branch, leaving the others inactive. The
selected branch applies a fixed detector-prediction set and a
deterministic local decision:
\[
\hat{y}=g_{\hat{z}}\left(
  \{d_m(x): m \in \mathcal{M}_{\hat{z}}\}
\right),
\]
where \(d_m(x)\in\{\mathrm{real},\mathrm{fake}\}\) is the thresholded
prediction from detector \(m\), and \(\mathcal{M}_{\hat{z}}\) is the
detector set assigned to the recovered domain. Thus, the router
recovers the missing audio type, while the selected branch combines the
predictions of its assigned expert or experts to produce \(\hat{y}\).

Detector outputs are therefore interpreted in a domain-conditioned score
space. A score reliable for speech may carry different decision
semantics for sound, singing, or music, despite the shared binary label
space. The system consequently recovers the domain condition before
applying the branch-local decision mapping, as shown in
Figure~\ref{fig:system_overview}.

\begin{figure*}[t!]
  \centering
  \includegraphics[width=\textwidth]{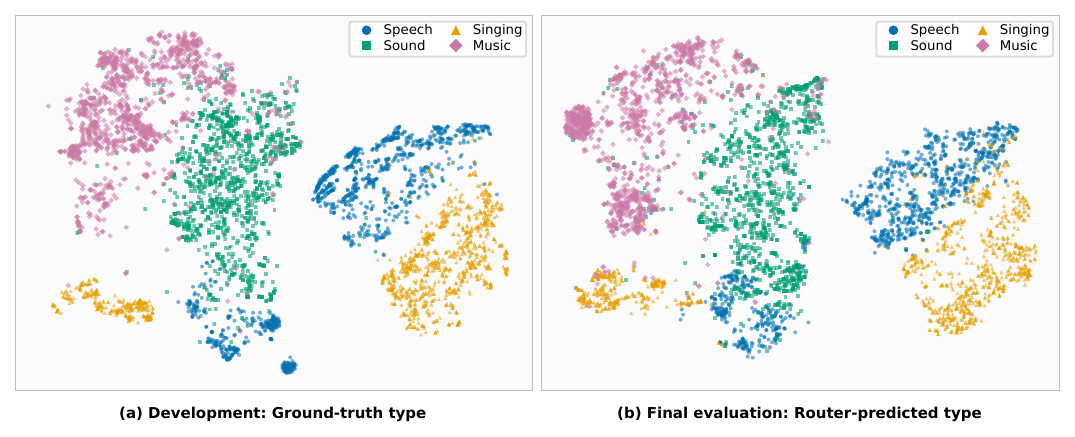}
  \caption{Router-domain representation geometry for AT-ADD Track2.
  Both panels use one t-SNE projection of balanced samples; development
  uses ground-truth types and final evaluation uses router-predicted
  types.}
  \Description{A two-panel scatter plot of BEATs 6-second router
  embeddings. The left panel shows development samples colored by
  ground-truth speech, sound, singing, and music labels. The right panel
  shows final-evaluation samples colored by router-predicted speech,
  sound, singing, and music labels.}
  \label{fig:embedding_geometry}
\end{figure*}

\subsection{Audio-Type Router}

The AudioType-BEATs-6s Router is a BEATs-based four-class classifier
with a 6-second input window. It predicts speech, sound, singing, or
music and is trained to recover the hidden audio type. Its output
provides the domain condition used to select the branch in which
downstream authenticity scores are interpreted.

The top-1 predicted type is used solely to select one of the four
decision branches. The router probability vector is not used in the
subsequent authenticity decision. Each selected branch instead applies
its predefined branch-local rule to the predictions of its assigned
expert or experts.

Real and fake examples appear in all four domains, but detector-relevant
cues and detector-score distributions can differ substantially across
domains. The predicted type selects the branch-specific score space used
for authenticity detection.

\subsection{Branch Detectors}

After routing, each branch selects a primary detector associated with
the recovered audio domain. Detector outputs are interpreted as
branch-local scores rather than domain-independent real/fake
posteriors. An auxiliary expert may additionally contribute during the
final fusion stage; this does not create a second routed branch.

The speech branch uses the Speech-XLSR Expert, which combines an XLS-R
frontend with an AASIST-style SSL countermeasure. Its speech-oriented
representation is used for speech-domain score interpretation.

The sound and music branches select the SoundMusic-EAT Expert as their
primary detector. During final fusion, its prediction is combined with
an auxiliary prediction from the Singing-EAT Expert using an OR-fake
rule: the fused prediction is fake if either expert predicts fake, and
real only if both experts predict real.

The singing branch uses the Singing-EAT Expert with test-time
aggregation. Its EAT-based representation is used for singing-domain
score interpretation, with temporal aggregation applied across crops.

\subsection{Final Decision}

The final decision function \(g_{\hat{z}}\) fuses the detector
predictions required by the selected branch. Decision thresholds are
selected separately for each audio-type branch by maximizing the
corresponding type-wise Macro-F1 on the development subset for that
audio type. Thus, the speech, sound, singing, and music decision rules
use thresholds estimated from the speech, sound, singing, and music
development subsets, respectively.

The final branch rules are deterministic. In the speech branch, five
temporal crops are evaluated independently with the Speech-XLSR Expert
using the speech-specific threshold. The final prediction is real only
when all five crops are predicted as real; if any crop is predicted as
fake, the final prediction is fake. The singing branch applies the same
five-crop conjunctive rule using the Singing-EAT Expert and the
singing-specific threshold. For the sound and music branches, the final
fusion stage combines the routed SoundMusic-EAT prediction with an
auxiliary Singing-EAT prediction. The two predictions use thresholds
selected on the corresponding audio-type development subset and follow
the OR-fake rule: the final prediction is fake if either expert predicts
fake and real only when both experts predict real.

\begin{table}[t!]
  \centering
  \caption{Development-set audio-type router comparison using a fixed
  6-second input window. Macro-F1 is computed across the four audio-type
  classes.}
  \label{tab:router_selection}
  \begin{tabular}{@{}lcc@{}}
    \toprule
    Representation & Acc. (\%) & Macro-F1 (\%) \\
    \midrule
    W2V2/XLS-R & 97.91 & 95.78 \\
    W2V-BERT & 97.58 & 95.71 \\
    EAT & 98.92 & 98.02 \\
    \textbf{BEATs} & \textbf{99.20} & \textbf{98.43} \\
    \bottomrule
  \end{tabular}
\end{table}

\section{Experiments and Results}

\subsection{Dataset and Evaluation Protocol}

We evaluate the system on AT-ADD Track2, an all-type audio deepfake
detection benchmark. Each submitted prediction is a binary real/fake
label, and the audio type is hidden at inference time. For each audio
type \(t\), the official evaluator first computes the Macro-F1 over the
real and fake classes:
\[
\operatorname{MacroF1}_{t}
=
\frac{1}{2}
\left(
F1_{t,\mathrm{real}} + F1_{t,\mathrm{fake}}
\right).
\]
The final Track2 score is the unweighted average over the four audio
types:
\[
\operatorname{MacroF1}_{\mathrm{T2}}
=
\frac{1}{4}
\sum_{t \in \mathcal{Z}}
\operatorname{MacroF1}_{t},
\]
where
\(\mathcal{Z}=\{\mathrm{speech},\mathrm{sound},\mathrm{singing},
\mathrm{music}\}\). We refer to the first quantity as type-wise
Macro-F1 and to the second as Track2 Macro-F1.

The official training and development splits provide both authenticity
labels and audio-type labels. Table~\ref{tab:track2_distribution}
summarizes sample counts and mean durations. The splits are imbalanced
with respect to both audio type and binary label: speech accounts for a
substantially larger portion of the data than music, and fake samples
are more frequent than real samples. This distribution makes type-wise
evaluation important, since aggregate performance can obscure weak
performance on smaller or acoustically distinct domains.

We use the closed Track2 setting. Training mixtures, crop durations,
checkpoint selection, and branch-local decision cutoffs are fixed
before evaluation using only the official training and development
resources. The official evaluation contains a Progress subset and the
full Eval set: Progress is an official feedback subset of Eval, while
the final leaderboard is computed on the full Eval set. The evaluation
data are blind at the sample level: participants submit binary
predictions, and the organizers return aggregate scores.
Development-set labels are therefore used for representation and
score-interpretation analysis, whereas final-set claims are based on
the official aggregate scores.

\subsection{Development Geometry and Router Selection}
\label{sec:representation_geometry}

We first examine whether the router representation captures audio-domain
structure. Figure~\ref{fig:embedding_geometry} visualizes balanced
development and final-evaluation samples in a shared t-SNE space
computed from AudioType-BEATs-6s Router embeddings. The development
panel is colored by ground-truth audio type, while the final-evaluation
panel is colored by router-predicted audio type because per-sample final
labels are not released. Development samples form type-dependent
regions. The final-evaluation panel visualizes the partition induced by
the deployed router; because ground-truth final audio types are
unavailable, it is not an independent validation of audio-type
separation. On the sampled PCA features used before t-SNE, the
silhouette score is 0.104 for development ground-truth type and 0.084
for final-evaluation predicted type.

The representation analysis supports treating audio type as a latent
condition for downstream score interpretation. The mean-duration
statistics in Table~\ref{tab:track2_distribution} further show that
speech and singing have mean durations around five seconds, whereas
sound and music are substantially longer. We use a 6-second router
window as a shared input window because it is close to the mean duration
of the vocal-domain clips while limiting reliance on longer sound and
music context. We then compare router representation families under the
same four-way audio-type recovery objective.

Table~\ref{tab:router_selection} reports router representation families
for hidden audio-type prediction. These metrics evaluate audio-type
recovery rather than real/fake detection under the fixed 6-second
router window. BEATs achieves the strongest type-recovery result and is
used as the router in the final routed detector.

\subsection{System Configuration}

\begin{table*}[t]
  \centering
  \caption{Official AT-ADD Track2 results and Progress-stage component
comparison (\%). Macro denotes Track2 Macro-F1, and the four domain
columns report type-wise Macro-F1. Standalone expert results are
available only for Progress.}
  \label{tab:official_final_results}
  \normalsize
  \setlength{\tabcolsep}{2.2pt}
  \begin{tabular}{@{}lrrrrrrrrrr@{}}
    \toprule
    & \multicolumn{5}{c}{Progress} & \multicolumn{5}{c}{Eval} \\
    \cmidrule(lr){2-6} \cmidrule(lr){7-11}
    System & Macro & Speech & Sound & Singing & Music
    & Macro & Speech & Sound & Singing & Music \\
    \midrule
    \multicolumn{11}{@{}l}{\textit{Official baselines}} \\
    Spec-ResNet
    & 53.22 & 51.08 & 52.92 & 48.41 & 60.48
    & 53.83 & 51.79 & 54.35 & 49.29 & 59.88 \\
    Qwen2.5-Omni-7B
    & 61.48 & 69.89 & 45.28 & 68.04 & 63.77
    & 61.78 & 69.29 & 45.94 & 68.03 & 63.87 \\
    AASIST
    & 62.38 & 63.69 & 56.88 & 64.08 & 64.87
    & 62.21 & 63.58 & 56.62 & 63.81 & 64.85 \\
    Qwen2.5-Omni-3B
    & 63.42 & 69.47 & 50.38 & 66.31 & 67.52
    & 63.23 & 68.74 & 50.41 & 65.78 & 67.99 \\
    WPT-XLSR-AASIST
    & 66.59 & 69.42 & 52.97 & 79.81 & 64.17
    & 66.68 & 69.31 & 53.83 & 79.56 & 64.04 \\
    FT-XLSR-AASIST
    & 79.25 & 79.43 & 66.08 & 96.33 & 75.17
    & 79.47 & 79.50 & 66.82 & 96.30 & 75.28 \\
    \midrule
    \multicolumn{11}{@{}l}{\textit{Standalone experts}} \\
    Speech-XLSR Expert
    & 81.43 & 86.41 & 83.86 & 89.58 & 65.86
    & \multicolumn{5}{c}{--} \\
    SoundMusic-EAT Expert
    & 92.66 & 79.79 & 97.70 & 94.17 & 98.99
    & \multicolumn{5}{c}{--} \\
    Singing-EAT Expert
    & 92.79 & 79.07 & 93.91 & 99.37 & 98.80
    & \multicolumn{5}{c}{--} \\
    \midrule
    \multicolumn{11}{@{}l}{\textit{Routed system}} \\
    \textbf{Ours}
    & \textbf{96.29} & \textbf{88.37} & \textbf{98.49}
    & \textbf{99.28} & \textbf{99.02}
    & \textbf{96.10} & \textbf{88.07} & \textbf{98.18}
    & \textbf{99.07} & \textbf{99.08} \\
    \bottomrule
  \end{tabular}
\end{table*}

The system consists of one audio-type router and three detector experts:
the Speech-XLSR Expert, the SoundMusic-EAT Expert, and the Singing-EAT
Expert. The router predicts the hidden audio type and selects the
primary expert branch. The final fusion stage then applies the
branch-specific aggregation rule, including auxiliary cross-expert
fusion for sound and music. All detector experts are trained on a single
L40S GPU.

\paragraph{AudioType-BEATs-6s Router.}
The router is a two-layer MLP over frozen BEATs embeddings. Audio longer
than 6 seconds is cropped to the first 6 seconds for embedding
extraction. The four-way classifier is trained from official
training-set audio-type labels with class-weighted cross-entropy. The
MLP head uses hidden dimension 256 and dropout 0.3, and is optimized
with Adam using learning rate \(10^{-3}\), weight decay \(10^{-4}\),
batch size 4096, and 15 epochs. The checkpoint with the highest
development-set Macro-F1 across the four audio-type classes is
selected.

\paragraph{Speech-XLSR Expert.}
The speech branch uses an XLS-R-300M frontend, an AASIST backend, and
4-second temporal crops. Training uses the official AT-ADD training
audio with codec augmentation, RawBoost, and SSI-style augmentation.
Additional signal-level transformations are applied to bona fide speech
to emulate TTS-like synthesis artifacts and replay-channel distortions,
and the resulting samples are used as augmented fake examples. The
model is optimized with Adam using learning rate
\(1.7136\times10^{-6}\), weight decay \(5\times10^{-4}\), batch size
24, and 20 epochs. Checkpoint selection uses development-set type-wise
Macro-F1 for the speech branch. At inference time, five temporal crops
are evaluated independently using the speech-specific decision
threshold. The speech-branch prediction is real only when all five
crops are predicted as real; if any crop is predicted as fake, the
final prediction is fake.

\paragraph{SoundMusic-EAT Expert.}
The SoundMusic-EAT Expert uses an EAT-large frontend and an AASIST
backend following the EAT-AASIST detector design
\cite{eataasist2026}. It is trained with an all-type mixture with
increased music sampling, codec augmentation, a binary authenticity
objective, and 10-second random crops. Inference uses a single
10-second score for the sound and music branches. Optimization uses
AdamW with a OneCycle schedule, frontend learning rate \(10^{-5}\),
backend learning rate \(10^{-4}\), weight decay \(10^{-2}\), batch
size 14, and up to 50 epochs. Early stopping and checkpoint selection
use development-set Track2 Macro-F1.

\paragraph{Singing-EAT Expert.}
The Singing-EAT Expert shares the EAT-large frontend and AASIST backend
with the SoundMusic-EAT Expert, but uses a non-speech mixture balanced
across sound, singing, and music. Training uses codec augmentation, a
binary authenticity objective, and 10-second random crops. At inference
time, five temporal crops are evaluated independently using the
singing-specific decision threshold. The singing-branch prediction is
real only when all five crops are predicted as real; if any crop is
predicted as fake, the final prediction is fake. The optimizer,
scheduler, and checkpoint selection settings follow those of the
SoundMusic-EAT Expert.

\subsection{Official Results and Analysis}

Table~\ref{tab:official_final_results} compares the routed system with
official baselines and Progress-stage component results. Since the
Progress set is a subset of Eval, Progress and Eval metrics are shown
side by side. Standalone experts are included only as Progress-stage
component results because Eval scores are unavailable for these runs.
The strongest official baseline, FT-XLSR-AASIST, obtains 79.25\%
Track2 Macro-F1 on Progress and 79.47\% on Eval. The routed system
obtains 96.29\% on Progress and ranks first on the final leaderboard
with 96.10\%.

On the Progress subset, the standalone experts show domain-dependent
specialization: Speech-XLSR performs best on speech, SoundMusic-EAT on
sound and music, and Singing-EAT on singing. The routed system selects
among these complementary branches, improving over the best standalone
expert by 3.50 percentage points.

The Progress-stage trend is consistent with Eval, where the routed
system scores only 0.19 percentage points lower. Relative to
FT-XLSR-AASIST, the final Track2 Macro-F1 gain is 16.63 percentage
points. The type-wise gains are 8.57, 31.36, 2.77, and 23.80 points for
speech, sound, singing, and music, respectively. The largest margins on
sound and music indicate the value of general-audio experts for
non-speech domains, whereas the smaller singing gain reflects the
already strong baseline on that domain.

Speech remains the primary limitation: its 88.07\% Eval type-wise
Macro-F1 is substantially below the three non-speech scores. This gap
motivates stronger speech-domain modeling and more reliable
branch-local score interpretation. It also demonstrates why type-wise
reporting is necessary: Track2 Macro-F1 summarizes overall performance
but does not expose which acoustic domain remains the bottleneck.

\subsection{Limitations}

The final evaluation set is blind at the sample level, and the
organizers release aggregate scores rather than per-sample final labels.
Analysis of the final set is therefore limited to aggregate-score
benchmark comparison. Confusion-matrix, generator-level, and per-error
attribution analyses require access to per-sample labels.

Our analysis draws on router representation geometry, router-family
probes, component-level Progress results, and official aggregate scores
on the final set. These sources support domain-conditioned score
interpretation under a hidden audio-domain condition.

\section{Conclusion}

We study the AT-ADD Track2 task as all-type audio deepfake detection
under a hidden audio-domain condition. Development-set geometry,
duration statistics, router-family probes, and component-level results
show that audio type is a key factor in detector-score interpretation.
The proposed system therefore first recovers the hidden audio domain
and then applies branch-local score interpretation, treating speech,
environmental sound, singing voice, and music as domains with distinct
score semantics.

The routed system implements this formulation with a BEATs-based
audio-type router, domain-specific detector experts, and branch-local
score interpretation. On the official final evaluation, it achieves
96.10\% Track2 Macro-F1 and ranks first on the AT-ADD Track2 final
leaderboard, with type-wise Macro-F1 scores of 88.07\%, 98.18\%,
99.07\%, and 99.08\% for speech, sound, singing, and music,
respectively. The type-wise results show strong performance in the
non-speech domains, while speech remains the main performance
bottleneck. These findings highlight the importance of
domain-conditioned representation and branch-local score interpretation
for all-type audio deepfake detection.

\bibliographystyle{ACM-Reference-Format}
\balance
\bibliography{references}

\end{document}